\documentclass[final]{aipproc}
\layoutstyle{6s}
\begin{document}
\title{GRB 970228 Within the EMBH Model}
\author{A. Corsi}{address={ICRA - International Center for Relativistic Astrophysics and Dipartimento di Fisica, Universit\`a di Roma ``La Sapienza'', Piazzale Aldo Moro 5, I-00185 Roma, Italy.}}
\author{M.G. Bernardini}{  address={ICRA - International Center for Relativistic Astrophysics and Dipartimento di Fisica, Universit\`a di Roma ``La Sapienza'', Piazzale Aldo Moro 5, I-00185 Roma, Italy.}}
\author{C.L. Bianco}{  address={ICRA - International Center for Relativistic Astrophysics and Dipartimento di Fisica, Universit\`a di Roma ``La Sapienza'', Piazzale Aldo Moro 5, I-00185 Roma, Italy.}}
\author{P. Chardonnet}{  address={Universit\'e de Savoie, LAPTH - LAPP, BP 110, F-74941 Annecy-le-Vieux Cedex, France.}}
\author{F. Fraschetti}{  address={Universit\`a di Trento, Via Sommarive 14, I-38050 Povo (Trento), Italy.}}
\author{R. Ruffini}{ address={ICRA - International Center for Relativistic Astrophysics and Dipartimento di Fisica, Universit\`a di Roma ``La Sapienza'', Piazzale Aldo Moro 5, I-00185 Roma, Italy.}}
\author{S.-S. Xue}{  address={ICRA - International Center for Relativistic Astrophysics and Dipartimento di Fisica, Universit\`a di Roma ``La Sapienza'', Piazzale Aldo Moro 5, I-00185 Roma, Italy.}}

\begin{abstract}We consider the gamma-ray burst of 1997 February 28 (GRB 970228) within the ElectroMagnetic Black Hole (EMBH) model. We first determine the value of the two free parameters that characterize energetically the GRB phenomenon in the EMBH model, that is to say the dyadosphere energy, $E_{dya}=5.1\times10^{52}$~ergs, and the baryonic remnant mass $M_{B}$ in units of $E_{dya}$, $B=M_{B}c^{2}/E_{dya}=3.0\times10^{-3}$. Having in this way estimated the  energy emitted during the beam-target phase, we evaluate the role of the InterStellar Medium (ISM) number density (n$_{ISM}$) and of the ratio ${\cal R}$ between the effective emitting area and the total surface area of the GRB source, in reproducing the observed profiles of the GRB 970228 prompt emission and X-ray (2-10~keV energy band) afterglow. The importance of the ISM distribution three-dimensional treatment around the central black hole is also stressed in this analysis.
\end{abstract}

\maketitle

The GRB 970228 \cite{IAUCCostaA} had an important role in solving the origin of GRBs through the first detection of counterparts at other wavelengths: the afterglow phenomenon, long-lived multi-wavelength emission, was discovered following GRB 970228 at X-ray (\cite{IAUCCostaB}, \citet{Costa}) and optical (\cite{IAUCGroot}, \citet{vanParadijs}) wavelengths. We consider of great interest to compare the predictions of the ElectroMagnetic Black Hole (EMBH) theory (see \citet{Brasile} and references therein) with the first afterglow observed by the Beppo-SAX satellite. We are also interested in testing the efficiency of the model in reproducing the GRB 970228 prompt emission: in the 40-700~keV energy band the burst was characterized by an initial $5$~s strong pulse followed, after about $30$~s, by three additional pulses of decreasing intensity (\citet{Frontera}). The InterStellar Medium (ISM) number density (n$_{ISM}$) inhomogeneities have an important role in interpreting this profile within the EMBH model.

Our analysis starts establishing the value of the two free parameters that determine energetically the GRB phenomenon in the EMBH model: the total energy deposited in the dyadosphere $E_{dya}$ (\citet{Rsw2000}) and the amount of the baryonic matter left over in the collapse process of the EMBH progenitor star (\citet{Rsw2000}), that can be parametrized by the dimensionless parameter $B=M_{B}c^{2}/E_{dya}$. With the choice of $E_{dya}=5.1\times10^{52}$~ergs and $B=3.0\times10^{-3}$, the EMBH model predicts that a 98\% of the total energy $E_{dya}$ is emitted during the so-called beam-target phase (\citet{lett2}), that is to say during the collision of the Accelerated Baryonic Matter-pulse (ABM-pulse) with the ISM (\citet{Brasile}). During this phase, the internal energy developed in the collision is instantaneously radiated away (fully radiative condition) and, as a consequence, the resulting shape of the light curve is strictly linked to the ISM distribution and number density (\citet{rbcfx02_letter}). We use a one-dimensional treatment of the ISM, where the n$_{ISM}$ is a function of the radial distance from the central black hole (\citet{rbcfx02_letter}). In order to reproduce the observed profile of the GRB 970228, n$_{ISM}$ has to range between the values of 10$^{-2}$~particles/cm$^{3}$ and $200$~particles/cm$^{3}$ in the region of space within $2.00\times10^{15}$~cm and $4.95\times10^{16}$~cm from the central black hole. Since $2.00\times10^{15}$~cm and beyond $4.95\times10^{16}$~cm, the ISM number density has a constant value of 1~particle/cm$^{3}$ (details are given in \citet{lungo}, \citet{MGX}). The correct spectral distribution of the energy emitted during the the beam-target phase depends on the ${\cal R}$ parameter (\citet{Spectr1}). As a consequence, the theoretical curves in selected energy bands are strictly related to this parameter. ${\cal R}$ is a function of the radial distance from the EMBH and it represents the ratio between the effective emitting area of the ABM-pulse and its total surface area:
\begin{equation}
{\cal R}=A_{eff}/A_{ABM}
\end{equation}
According to \citet{Spectr1}, by assuming a black-body spectrum in the co-moving frame for the radiation emitted during the collision with the ISM, the spectral distribution of the energy emitted results to be dependent on the temperature of the emitting black body (\citet{Spectr1}, \citet{cospar02}):
\begin{equation}
T=\left(\frac{\Delta E_{int}}{4\pi r^{2}\Delta \tau \sigma {\cal R}}\right)^{1/4}
\label{eq temp}	
\end{equation}
where $\Delta E_{int}$ is the proper internal energy developed in the collision of the ABM-pulse with the ISM in the proper time interval $\Delta \tau$, $r$ is the radial coordinate of the ABM-pulse, $t$ is the laboratory time (\citet{lett1}), $\sigma$ is the Stefan-Boltzmann constant. In the case of GRB 970228 we find ${\cal R}$ monotonically varying from $3.7\times10^{-12}$ to $8.8\times10^{-11}$ when the radial coordinate $r$ goes from $7.0\times10^{14}$ cm to $5.0\times10^{17}$ cm. With this result, the first peak in the 40-700~keV observed light curve is correctly reproduced by the model (details are given in \citet{lungo}, \citet{MGX}). The three additional pulses, that follow the first one after a gap in the emission, are reproduced by the model in terms of the mean luminosity. The Fast Rise Exponential Decay (FRED) shape that emerges in the theoretical light curve is a consequence of the one-dimensional treatment of the ISM. To solve this problem, a three-dimensional treatment of the ISM distribution is required (details are given in \citet{lungo}, \citet{MGX}). 

\begin{figure}	
\centering
\includegraphics[width=\hsize]{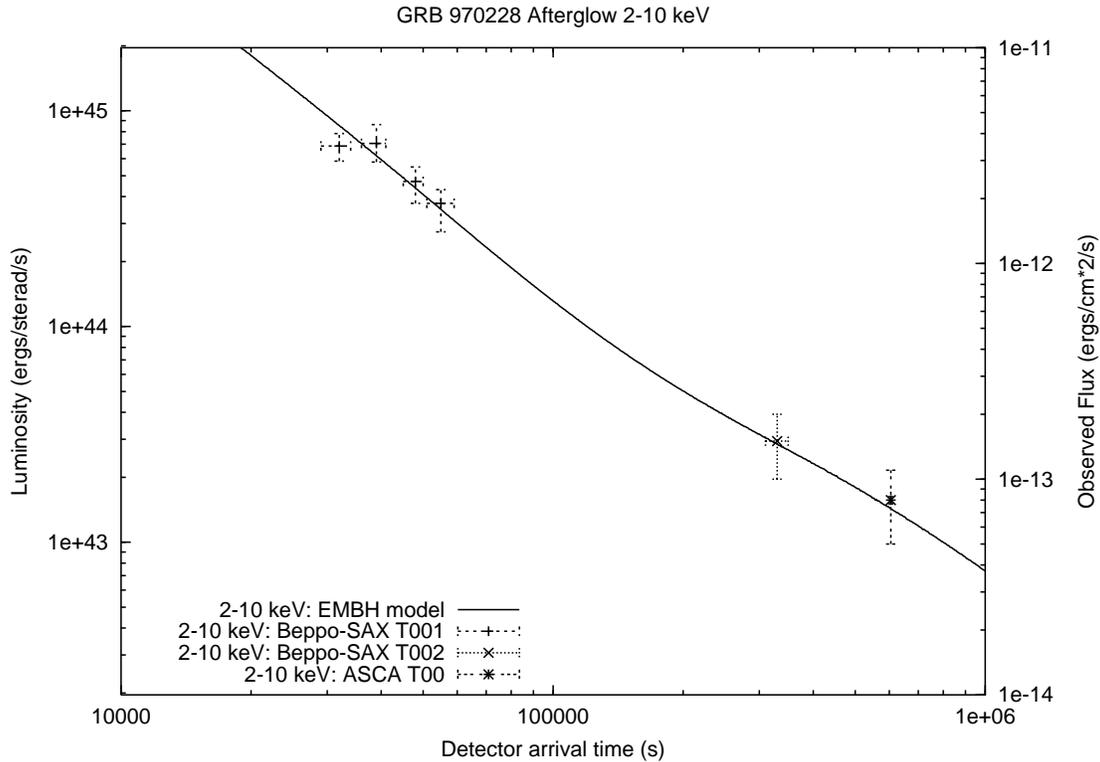}
\caption{Afterglow 2-10 keV: the solid line represents the theoretical light curve for the 2-10 keV emission in the EMBH model. The points are the GRB 970228 2-10 keV afterglow data observed by Beppo-SAX (\citet{Costa}) and ASCA (\cite{IAUC})}.	
\label{fig:afterglow2-10}
\end{figure}

About the X-ray afterglow, in Fig.\ref{fig:afterglow2-10} we present the theoretical curve in the 2-10~keV energy band compared with the observed data by Beppo-SAX (\citet{Costa}) and ASCA \cite{IAUC}. The afterglow phase corresponds to the ABM-pulse expansion in the region beyond $4.95\times10^{16}$~cm, where the number density of the ISM has a constant value, n$_{ISM}=$1particle/cm$^{3}$. We can see that there is a good agreement ($\chi^{2}$=0.5) between the theoretical light curve in the 2-10~keV energy band and the observed data by Beppo-SAX and ASCA.
From this analysis we conclude that:
\begin{itemize}	
\item a mask of density inhomogeneities of the ISM is needed in the region of space between $2.00\times10^{15}$~cm and $4.95\times10^{16}$~cm from the black hole, in order to reproduce the structure of the GRB 970228 prompt emission; 	\item a three-dimensional treatment of the ISM is required in order to improve the theoretical predictions of the model (details are given in \citet{lungo}, \citet{MGX});
\item finally, a good result is obtained with a constant value of the n$_{ISM}=$1particle/cm$^{3}$ for the 2-10~keV afterglow emission.
\end{itemize}

\bibliographystyle{aipproc}
\bibliography{corsi_alessandra_0}

\end{document}